\documentclass{article}
\newcommand{\be}{\begin{equation}}
\newcommand{\ee}{\end{equation}}

\begin{document}

\title {\bf{An Exact Solution for Static Scalar Fields Coupled to Gravity in
$(2+1)$-Dimensions}}

\author{\bf{ Durmus Daghan and Ayse H. Bilge}}

\maketitle \noindent Department of Mathematics, Faculty of Science
and Letters,
Istanbul Technical University, Maslak, 80626 Istanbul, Turkey. daghand@itu.edu.tr, bilge@itu.edu.tr\\
\\

\hspace*{35pt}------------------------------------------------------------------------------------
\\
\hspace*{50pt}\noindent {\bf{Abstract.}} We obtain an exact
solution for the Einstein's \hspace*{50pt}equations with
cosmological constant coupled  to a scalar, static
\hspace*{50pt}particle in static, "spherically" symmetric
background in $2+1$ \hspace*{50pt}dimensions.

\hspace*{35pt}------------------------------------------------------------------------------------
\hspace*{50pt}\noindent KEY WORDS: black holes; Choptuik
formation; singularity.
\\

\section*{{1. Introduction}}\setcounter{section}{1}\mbox{}
In reference [1], Choptuik studied numerically a massless scalar
field $\phi$ minimally coupled to the gravitational metric in
$3+1$ dimensions and found a scaling relation
$M=C(p-p_{*})^{\gamma}$ for the black hole mass $M$ in the limit
$p$ close to critical value $p_{*}$ where $\gamma$ is
approximately equal to 0.374 for all 1-parameter families of
scalar data. A similar behavior was also obtained for the
cylindrically symmetric case in $3+1$ dimensions studied by
Abrahams and Evans [2].

In references 3-6, similar systems in $2+1$ dimensions were
studied numerically and analytically. In reference [7], Birkandan
and Hortacsu studied the BTZ system along the lines of Pretorius
and Choptuik [3] work. They dropped the time dependency and first
studied the static case with no scalar field and then, added the
scalar field perturbatively.

In this note we study the Einstein Klein-Gordon system in AdS
spacetime for the static spherically symmetric background. The
original equations studied perturbatively in [7] are reduced to a
second order system and then, the exact solution are obtained in a
new coordinate frame. Explicit solutions in terms of the first
coordinate frame involve hypergeometric functions they were
omitted. We used the further coordinate transformations to obtain
the exact solution in the form generalizing the AdS metric as
given by Matschull [10].

\section*{{2.  Einstein's Field Equations}}
\setcounter{section}{2} \setcounter{equation}{0}\mbox{}

Let $M$ be a 3-dimensional Lorentzian manifold with local
coordinates $(t,r,\theta)$ and $g_{\mu\nu}$ be a metric on $M$. We
start with the corresponding line element

\be ds^{2}= -e^{2X(r)}dt^{2}+e^{2X(r)} dr^2+e^{-2Y(r)}d\theta^2,
\ee We shall use the notation $X'=\frac{dX}{dr}$ and
$Y'=\frac{dY}{dr}$. We give below the non-zero components of the
Christoffel symbols of the second kind \be \Gamma^{r}_{tt}=
X',\quad \Gamma^{t}_{tr}= X',\quad \Gamma^{r}_{rr}= X',\quad
\Gamma^{\theta}_{r \theta}=-Y',\quad \Gamma^{r}_{\theta \theta}=
Y'e^{-2(X+Y)}, \ee of the  curvature tensor \be
R_{trtr}=e^{2X}X'',\hspace*{3pt} R_{t \theta t
\theta}=e^{-2Y}(-X'Y'),\hspace*{3pt} R_{r \theta r
\theta}=e^{-2Y}(Y''-X'Y'-Y'^2), \ee of the Ricci tensor \be
R_{tt}=
X''-X'Y',\hspace*{3pt}R_{rr}=-X''+Y''-X'Y'-Y'^{2},\hspace*{3pt}
R_{\theta\theta}=e^{-2(X+Y)}(Y''-Y'^{2}), \ee and the Ricci scalar
$R$ \be R=2e^{-2X}(-X''+Y''-Y'^{2}). \ee The Einstein's equations
with cosmological constant coupled to a scalar, static particle
are [8] \be R_{\mu\nu}-\frac{1}{2}g_{\mu\nu}R+ \Lambda
g_{\mu\nu}=\kappa T_{\mu\nu}, \ee where \be
T_{\mu\nu}=\partial_{\mu}\phi
\partial_{\nu}\phi- \frac{1}{2}g_{\mu\nu} (\partial_{\lambda}\phi
\partial^{\lambda}\phi),
\ee and $\phi$ satisfies the wave equation

\be g^{\mu\nu} \nabla_{\mu} \nabla_{\nu} \phi =
(-g)^{-\frac{1}{2}}\partial_{\mu}[(-g)^{\frac{1}{2}}g^{\mu\nu}
\partial_{\nu}\phi],
\ee with $\det\vert g_{\mu\nu}\vert=g$ [9]. From Eqn.(7), we
obtain

\be T_{tt}=\frac{1}{2}\phi'^{2},\quad
T_{rr}=\frac{1}{2}\phi'^{2},\quad
T_{\theta\theta}=-\frac{1}{2}e^{-2(X+Y)}\phi'^{2}. \ee The wave
equation for the scalar field is reduced to

\be \phi''-Y'\phi'=0 \ee which can be integrated as \be
\phi'=\frac{\lambda}{\sqrt{2\pi}}e^{Y}, \ee where $\lambda$ is an
integration constant.

Following the conventions in [3] we choose $\kappa=4\pi$ and after
relevant substitutions, (6) reduces to the following system of
ODE's for $X$, $Y$ and $\phi$,
\begin{eqnarray}
X''+\Lambda e^{2X}+ \lambda^2 e^{2Y}   =0,\cr
 X'Y'-\Lambda e^{2X}+\lambda^2 e^{2Y}=0 ,\cr
 Y''-X'Y'-Y'^{2}-\Lambda  e^{2X}-\lambda^2 e^{2Y}=0 .\cr
\end{eqnarray}
\vskip 0.1cm \noindent {\bf Remark 1.} If we substitute $Y=X$ in
the system (12), it can be seen that after algebraic eliminations
the third equation gives $-4\Lambda e^{2X}=0$, hence $\Lambda$
should be zero for compatibility. On the other hand for $Y\ne X$,
(12) is a consistent system. In the following we
 shall assume that $Y\ne X$.
\vskip 0.3 cm

We now proceed with the solution of the system (12). We shall
first reduce it to a second order system then obtain an analytic
solution in a suitable coordinate system.

\vskip 0.3cm \noindent \textit{\textbf{Proposition}} {\bf 2.}
\textit{The system of ODE's (12) can be reduced to the following
system of ODE's for $X,Y$}

\begin{eqnarray}
X'= \frac{1}{2}c e^{Y}\mp \frac{1}{2}[c^{2}e^{2Y}+4(-\Lambda
e^{2X} +\lambda^{2}e^{2Y})]^{\frac{1}{2}} \cr Y'= \frac{1}{2}c
e^{Y}\pm \frac{1}{2}[c^{2}e^{2Y}+4(-\Lambda e^{2X}
+\lambda^{2}e^{2Y})]^{\frac{1}{2}}
\end{eqnarray}
\textit{where  $c$ and $\lambda$ are integration constants and
$\Lambda$
is the cosmological constant.}\\

\noindent {\bf Proof.} After algebraic eliminations, the system
(12) is reduced to
\begin{eqnarray}
X''+\Lambda e^{2X}+\lambda^{2}e^{2Y}=0,\cr X'Y'-\Lambda
e^{2X}+\lambda^{2}e^{2Y}=0,\cr Y''=Y'^{2}+2\Lambda e^{2X}.
\end{eqnarray}
Let  $Z=X+Y$ and differentiate twice with respect to $r$ to get
$Z''=X''+Y''$. Substituting the expressions of $X''$ and $Y''$ we
obtain

\be Z''=\Lambda e^{2X}-\lambda^{2}e^{2Y}+Y'^{2}
   =X'Y'+Y'^{2}=Z'Y'.
\ee Note that $Z'=0$ is a special solution. For $Z'\neq 0$, we can
integrate Eqn.(15) as $Z'=c e^{Y}$ where $c$ is an integration
constant. Hence the system (14) gives
\begin{eqnarray}
 X'+Y'=c e^{Y},\cr
    X'Y'-\Lambda e^{2X}+\lambda^{2}e^{2Y}=0.
\end{eqnarray}
Solving $X'$ from the first and substituting in the second, we
obtain
$$(c e^{Y}-Y')Y'-\Lambda e^{2X}+\lambda^{2}e^{2Y}=0.$$
Rearranging this equation and completing the square, we have
$$(Y'-\frac{1}{2}ce^{Y})^{2}=-\Lambda
e^{2X}+\lambda^{2}e^{2Y}+\frac{1}{4}c^{2}e^{2Y}.$$ Taking the
square root and rearranging again we obtain $Y'$ and $X'$ as in
(13) hence the proof is complete.\hfill $\bullet$\\

We note that the special solution $Z'=X'+Y'=0$ corresponds to the
value $c=0$ of the integration constant.

\vskip 0.3cm \noindent {\bf Remark 3.} For $X'+Y'=0$, $c=0$ and
the metric reduces to the form $$ ds^{2}=e^{2X}\left( -dt^{2}+
dr^2+e^{-2Y_o}d\theta^2\right),$$ where $Y_0$ is an integration
constant. Unless otherwise indicated, we shall assume that $c\ne
0$. We note that the solutions corresponding to different choices
of the sign of the square root in (13) are related by the sign
changes of $c$ and $r$, hence we take
\begin{eqnarray} X'= \frac{1}{2}c e^{Y}-
\frac{1}{2}[c^{2}e^{2Y}+4(-\Lambda e^{2X}
+\lambda^{2}e^{2Y})]^{\frac{1}{2}}\hfill \cr Y'= \frac{1}{2}c
e^{Y}+ \frac{1}{2}[c^{2}e^{2Y}+4(-\Lambda e^{2X}
+\lambda^{2}e^{2Y})]^{\frac{1}{2}},\quad c\ne 0,
\end{eqnarray}
allowing $c$ to have both positive and negative signatures.
 \vskip
0.3cm

The metric (1) admits obviously the abelian ${\bf{g_2}}$ symmetry
algebra generated by $\frac{\partial}{\partial_t}$ and
$\frac{\partial}{\partial_\theta}.$ The computation of the
Killing's equation
$$\xi_{\mu;\nu}+\xi_{\nu;\mu}=0$$
is straightforward. It can be seen that for $c\neq0$ there are
only two Killing vectors, i.e.,

\be X'+Y'\ne0:\quad \xi^{(1)}=\partial_{t}, \quad
\xi^{(2)}=\partial_{\theta}. \ee For $c=0$ there is third Killing
vector and the symmetry algebra has the following generators.

\be X'+Y'=0:\quad \xi^{(1)}=\partial_{t}, \quad
\xi^{(2)}=\partial_{\theta},\quad \xi^{(3)}=\theta
\partial_{t} +t\partial_{\theta} \ee \vskip 0.3cm

Note that as the Ricci tensor is diagonal, using (4-5) and (14) we
can obtain the curvature invariants as
$$R_{t}^{t}=2\Lambda,\quad R_{r}^{r}=2\Lambda+2\lambda^2 e^{-2X+2Y},\quad
R_{\theta}^{\theta}=2\Lambda,\quad R=6\Lambda + 2\lambda^2
e^{-2X+2Y}.$$ These expressions show that curvature singularities
are  related either to the cosmological constant $\Lambda$  or to
the particle coupling manifested by $\lambda$. In the following we
shall assume that $\Lambda=-1$ and use a series of coordinate
transformations to obtain an exact analytic solution.

\vskip 0.3cm \noindent \textit{\textbf{Proposition}} {\bf 4.}
\textit{The system of ODE's (17) is equivalent to the system} \be
\rho'= \frac{1}{2}\rho^2 \left[ \mu \sin\varphi-
\cos2\varphi\right],\ee \be \varphi'= \rho \ \sin\varphi\
\cos\varphi,\ee \textit {where $\rho$ and $\varphi$ are defined by
}$$e^X=\frac{1}{2}\rho\ \ \cos\varphi,\quad
e^Y=\frac{1}{\sqrt{c^2+4\lambda^2}}\rho\ \ \sin\varphi,$$
\textit{and  $\mu=\frac{c}{\sqrt{c^2+4 \lambda^2}}$.}
\\

\noindent {\bf Proof.} Taking derivatives of $e^X$ and $e^Y$ and
substituting in (17), we obtain
$$\rho' \ \cos\varphi -\varphi'\rho\  \sin\varphi
=\frac{1}{2}\rho^2 \left[\mu\  \sin\varphi\
\cos\varphi-\cos\varphi\right],$$
$$\rho' \ \sin\varphi +\varphi'\rho\ \cos\varphi=\frac{1}{2}\rho^2
\left[\mu \ \sin^2\varphi+\sin\varphi\right].$$ By multiplying the
first equation with  $\cos\varphi$ and the second one with
$\sin\varphi$ and adding, we obtain $\rho'$ as in (20). By a
similar elimination we obtain $\varphi'$, hence the proof is
complete.\hfill $\bullet$\\

We will now obtain $\rho$ as a function of $\varphi$, as it will
be discussed below, (20) can be seen as a parameter transformation
hence we have an analytic solution in the coordinate frame
$(t,\rho,\varphi)$. Explicit solutions in terms of $r$ involve
hypergeometric functions and shall be omitted.\\

\noindent \textit{\textbf{Proposition}} {\bf 5.} \textit{ The
solution of system of ODE's $(12)$ in the coordinate frame
$(t,\rho,\varphi)$, with $0<\varphi<\pi/2$ is \be
ds^2=-\left[\frac{1}{2}\rho\  \cos\varphi \right]^{2}
dt^{2}+\left[ \frac{1}{2} \frac{1}{\sin\varphi}\right]^{2} d
\varphi^2+ \left[ \frac{\sqrt{c^2+4\lambda^2}}{\rho\
\sin\varphi}\right]^2 d\theta^2 ,\ee where \be
\rho(\varphi)=\frac{\rho_{0}}{\sqrt{2}}
[1+\sin\varphi]^{\frac{\mu}{2}} [\sin\varphi]^{-\frac{1}{2}}
[\cos\varphi]^{\frac{-1-\mu}{2}} .\ee }

\noindent {\bf Proof.} From the Eqn.(20) and (21), we obtain
$$\frac{d\rho}{d\varphi}=\rho \left[\frac{1}{2} \frac{\mu}{\cos\varphi}-
\cot2\varphi\right]$$ which leads to
$$\frac{d\rho}{\rho}=\left[\frac{1}{2} \frac{\mu}{\cos\varphi}- \cot2\varphi\right]d\varphi.$$
Integrating both sides, we obtain $\rho$ as a function of
$\varphi$ as given Eqn.(23). As $e^X$ and $e^Y$ are positive, the
coordinate transformation used in Proposition 4 implies that
$0<\varphi<\pi/2$.

Substituting $\rho$ in $e^X$ and $e^Y$ and using
$dr=\frac{dr}{d\varphi}d\varphi$,
$dr=(\frac{d\varphi}{dr})^{-1}d\varphi$, we obtain
$$dr^2=\left[\frac{1}{\rho \sin\varphi\cos\varphi}
\right]^{2}d\varphi^2$$ and reach  Eqn.(22) after
rearrangements.\hfill $\bullet$\\

The expression of the curvature invariants are
$$R_{t}^{t}=-2,\quad R_{\varphi}^{\varphi}=-2+\frac{8\lambda^2}
{(c^2+4\lambda^2)}\tan^2\varphi,\quad R_{\theta}^{\theta}=-2,$$
$$\quad R=-6+\frac{8\lambda^2}{(c^2+4\lambda^2)}\tan^2\varphi.$$
Thus there is a curvature singularity at $\varphi=\frac{\pi}{2}$.
Comparing with the previous expressions of the curvature
invariants, it can be seen that the singularity at
$\varphi=\frac{\pi}{2}$ is related
to the scalar particle.\\

We shall now use further coordinate transformations to obtain the
exact solution (22) in the form generalizing the AdS metric as
given in [10]. Substituting for $\rho$, Eqn.(22) reduces to\\

$\hspace*{45pt} ds^2=-\frac{\rho_0^2}{8} [\sec\varphi+\tan \varphi
]^\mu \cot\varphi\ dt^2 +[\frac{1}{2\ \sin
\varphi}]^2\ d\varphi^2\\
\hspace*{87pt}+\frac{2(c^2+4\lambda^2)}{\rho_0^2}
[\sec\varphi+\tan\varphi ]^{-\mu} \cot\varphi\ d\theta^2.$\\

\noindent First let
$$\frac{d\varphi}{\sin\varphi}=dv.$$
Then $v=\ln(\tan \frac{\varphi}{2})$ or $e^v=\tan
\frac{\varphi}{2}$. Using double angle formulas for the tangent
function we can see that
$$\tan\varphi=\frac{2e^v}{1-e^{2v}}=-\frac{1}{{\rm sinh} v}.$$
As $\frac{\varphi}{2}$ varies from zero to $\frac{\pi}{4}$, then
$0<\varphi<\frac{\pi}{2}.$ $v$ is negative $(-\infty< v <0)$,
hence $${\rm sec}\hspace*{2pt}  \varphi=-\frac{{\rm cosh} v}{{\rm
sinh} v}.$$ Finally setting  $u= -v$ we obtain the metric
$$ds^2=-\frac{\rho_0^2}{8} \left[\frac{1+{\rm cosh} u } {{\rm sinh} u}\right]^\mu
{\rm sinh}u\ dt^2 + \frac{1}{4} \ du^2
+\frac{2(c^2+4\lambda^2)}{\rho_0^2} \left[\frac{1+{\rm cosh} u }
{{\rm sinh} u}\right]^{-\mu} {\rm sinh}u\ d\theta^2,
$$

\noindent where $0<u<\infty$. At the last step we set $u=2\chi$
and use double angle formulas to obtain \be
ds^2=-\frac{\rho_0^2}{4} \left[\frac{{\rm cosh}\chi}{{\rm
sinh}\chi}\right]^\mu
                {\rm cosh}\chi \ {\rm sinh}\chi \ dt^2
      +d\chi^2
+\frac{4} {\rho_0^2}(c^2+4\lambda^2) \left[\frac{{\rm
cosh}\chi}{{\rm sinh}\chi}\right]^{-\mu}
                {\rm cosh}\chi \ {\rm sinh}\chi \ d\theta^2. \ee

The coefficients $\frac{\rho_0^2}{4}$ and $\frac{4}{\rho_0^2}$ can
be eliminated by appropriate scalings of   $t$ and $\theta$ as
$$\tilde{t}=\frac{\rho_0}{2}t,\quad \tilde{\theta}=\frac{2}{\rho_0}\theta$$
and the metric is \be ds^2=-\left[\frac{{\rm cosh}\chi}{{\rm
sinh}\chi}\right]^\mu
                {\rm cosh}\chi \ {\rm sinh}\chi \ d{\tilde{t}}^2
      +d\chi^2
+(c^2+4\lambda^2) \left[\frac{{\rm cosh}\chi}{{\rm
sinh}\chi}\right]^{-\mu}
                {\rm cosh}\chi \ {\rm sinh}\chi \ d{\tilde\theta}^2.\ee

Note that $\sqrt{c^2+4\lambda^2}=\frac{c}{\mu}$ induces a scaling
of $\theta $ due to the effect of the coupled scalar field which
can be interpreted as inducing a topological defect. For
$\lambda=0$, $\mu=\frac{c}{\vert c\vert}=\pm 1$. Taking $c=1$ we
have AdS solution
$$ds^2=-{\rm cosh}^2\chi\ d{\tilde{t}}^2
      +d\chi^2
+{\rm sinh}^2\chi \ d{\tilde\theta}^2
$$
as given by Eqn.(1.1) in [10].\\

\section*{{3. Results and Further Discussions}}\mbox{}

We found an exact solution in $2+1$ dimensions for the Einstein's
equations with cosmological constant coupled  to a scalar, static
particle in static, spherically symmetric background which was
studied perturbatively in reference [7]. Our solution involves a
parameter $\mu$, $0<\mu<1$ related to strength of the coupling
field and AdS limit is reached for $\mu=1$.

Furthermore, in the final metric (25) the scaling
$\tilde{\theta}\rightarrow\frac{c}{\mu}\tilde{\theta}$ where $c$
is an integration constant can be interpreted as a topological
defect. Although, the absence of $t$ dependency rules out the
existence of trapped surfaces [11] page 435, the metric given in
(25) provides an exact module for the problem studied in Matschull
[10] with cutting and gluing procedure.

\section*{{Acknowledgements}}

The authors would like to thank Professors G. Clement and A.
Fabbri for pointing out that our solution as given by Eqn. 25, has
been obtained previously in reference [13], and also they would
like to thank Professor M. Hortacsu for illuminating discussions
and for pointing out reference [10].

\section*{{Note added in proof}}

The solution for a special case of the metric (1),  corresponding
to choice of $\lambda=e^{-Y}$ and $\nu(\lambda)=X$ is obtained in
reference 12.  The analytic solution to the complete problem
investigated here  has been previously obtained in reference 13.
With the signature $(+,-,-)$ our metric (25) coincides with
equation (3.7) in [13] as below. The equation (3.7) is

$\hspace*{30pt} ds^2=A\mid \rho-\rho_+\mid^{\frac{1}{2}+a}  \mid
\rho-\rho_-\mid^{\frac{1}{2}-a}
dt^2\\\hspace*{55pt}-\frac{4\mid\Lambda\mid}{A} \mid
\rho-\rho_+\mid^{\frac{1}{2}-a} \mid
\rho-\rho_-\mid^{\frac{1}{2}+a} d\theta^2
+\frac{1}{4\Lambda(\rho-\rho_+)(\rho-\rho_-)}d\rho^2.$\\

For $\Lambda=-1$, $a=-\frac{\mu}{2}$ and
$\rho=\frac{1}{2}\left[(\rho_+ -\rho_-) \cosh 2\chi+(\rho_+
+\rho_-)\right]$ it reduces to

$\hspace*{30pt} ds^2=A\mid \rho_+ -\rho_-\mid\hspace*{3pt}\mid
\cosh \chi\mid^{1+\mu} \hspace*{3pt} \mid \sinh \chi\mid^{1-\mu}
dt^2\\\hspace*{55pt}-\frac{4\mid \rho_+-\rho_-\mid}{A}\mid \cosh
\chi\mid^{1-\mu} \hspace*{3pt} \mid \sinh \chi\mid^{1+\mu}
d\theta^2
-d\chi^2$\\

\noindent where $A$ and $\mid \rho_+ -\rho_-\mid$ are constants.
Thus with appropriate scalings, we get equation (25).

\end{document}